\documentclass[twocolumn,showpacs,preprintnumbers,amsmath,amssymb]{revtex4}
\usepackage{graphicx}
\usepackage{dcolumn}
\usepackage{bm}
\usepackage{color}
\begin{document}

\newcommand{\kvec}{\mbox{{\scriptsize {\bf k}}}}
\newcommand{\lvec}{\mbox{{\scriptsize {\bf l}}}}
\newcommand{\qvec}{\mbox{{\scriptsize {\bf q}}}}
%
\def\eq#1{(\ref{#1})}
\def\fig#1{Fig.\hspace{1mm}\ref{#1}}
\def\tab#1{\hspace{1mm}\ref{#1}}
%
\title{Electron-phonon pairing mechanism: cuprates with high value of the critical temperature}
\author{R. Szcz{\c{e}}{\`s}niak, A.P. Durajski}
\email{adurajski@wip.pcz.pl}
\affiliation{Institute of Physics, Cz{\c{e}}stochowa University of Technology, Al. Armii Krajowej 19, 42-200 Cz{\c{e}}stochowa, Poland}
\date{\today}
\begin{abstract}
The model for the cuprates based on the modified electron-phonon pairing mechanism has been tested. For this purpose, the superconductors with high value of the critical temperature have been taken into consideration. In particular: ${\rm YBa_{2}Cu_{3}O_{7-y}}$, ${\rm HgBa_{2}CuO_{4+y}}$, 
${\rm HgBa_{2}Cu_{1-x}Zn_{x}O_{4+y}}$, and ${\rm HgBa_{2}Ca_{2}Cu_{3}O_{8+y}}$. It has been shown that the dependence of the ratio $R_{1}\equiv 2\Delta_{tot}^{\left(0\right)}/k_{B}T_{C}$ on the doping ($p$) can be properly predicted in the framework of the presented theory; the symbol $\Delta_{tot}^{\left(0\right)}$ denotes the energy gap amplitude at the temperature of zero Kelvin, and $T_{C}$ is the critical temperature. The numerical results have been supplemented by the formula which describes the function $R_{1}\left(p\right)$.
\end{abstract}
\maketitle

In the solid state physics, the issue of the correctly determining the pairing mechanism in the high-temperature superconductors (cuprates) is very controversial \cite{Bednorz1986A}. Essentially, there is a clash of the two views. First of them is based on the original Fr{\"o}hlich idea indicating that the interaction of the electron gas with the phonons is responsible for the formation of the superconducting state \cite{Frohlich1950A}. The second view favors the clean electron correlations, which in the simplest way can be modeled by the one-band Hubbard Hamiltonian \cite{Hubbard1963A}.

It is true that in the cuprates, the classical electron-phonon interaction is over one level weaker than the electron correlations \cite{Hauge2006A}. Hence, it may seem that the limitation to the correlations in the electronic subsystem is correct. Unfortunately, in the framework of the Hubbard model with the positive values of the potential ($U_{H}$) it is very difficult to describe the thermodynamic properties of the high-temperature superconducting state.   
What is even worse, there is no convincing evidence that the superconducting state can exist at the sufficiently high temperature \cite{Scalapino1992A}. On the other hand, in the case of the negative value of the interaction potential with the same intensity of $U_{H}$, proving the existence of the superconducting state with the high critical temperature is a simple matter.   

Additionally, it is easy to prove that the pairing mechanism based on the classical electron-phonon interaction is insufficient to fully describe the superconducting state in the cuprates \cite{Szczesniak2001A}, \cite{Szczesniak2006B}. 

In the presented paper, we have tested the pairing mechanism, which combines two opposing approaches. The new mechanism has been proposed recently in  \cite{Szczesniak2012D} (see also \cite{Szczesniak2012E} and \cite{Szczesniak2012F}). It is based on three postulates:
(i) {\it In the superconductivity domain of the cuprates the fundamental role is played by the electrons on the $CuO_{2}$ planes.}
(ii) {\it In the cuprates there exists the conventional electron-phonon interaction, which does not have to be strong.}
(iii) {\it In the cuprates there exist strong electronic correlations, but the electron-electron scattering in the superconductivity domain is inseparably connected with absorption or emission of the vibrational quanta.}

%
\begin{figure}[ht]
\includegraphics*[width=\columnwidth]{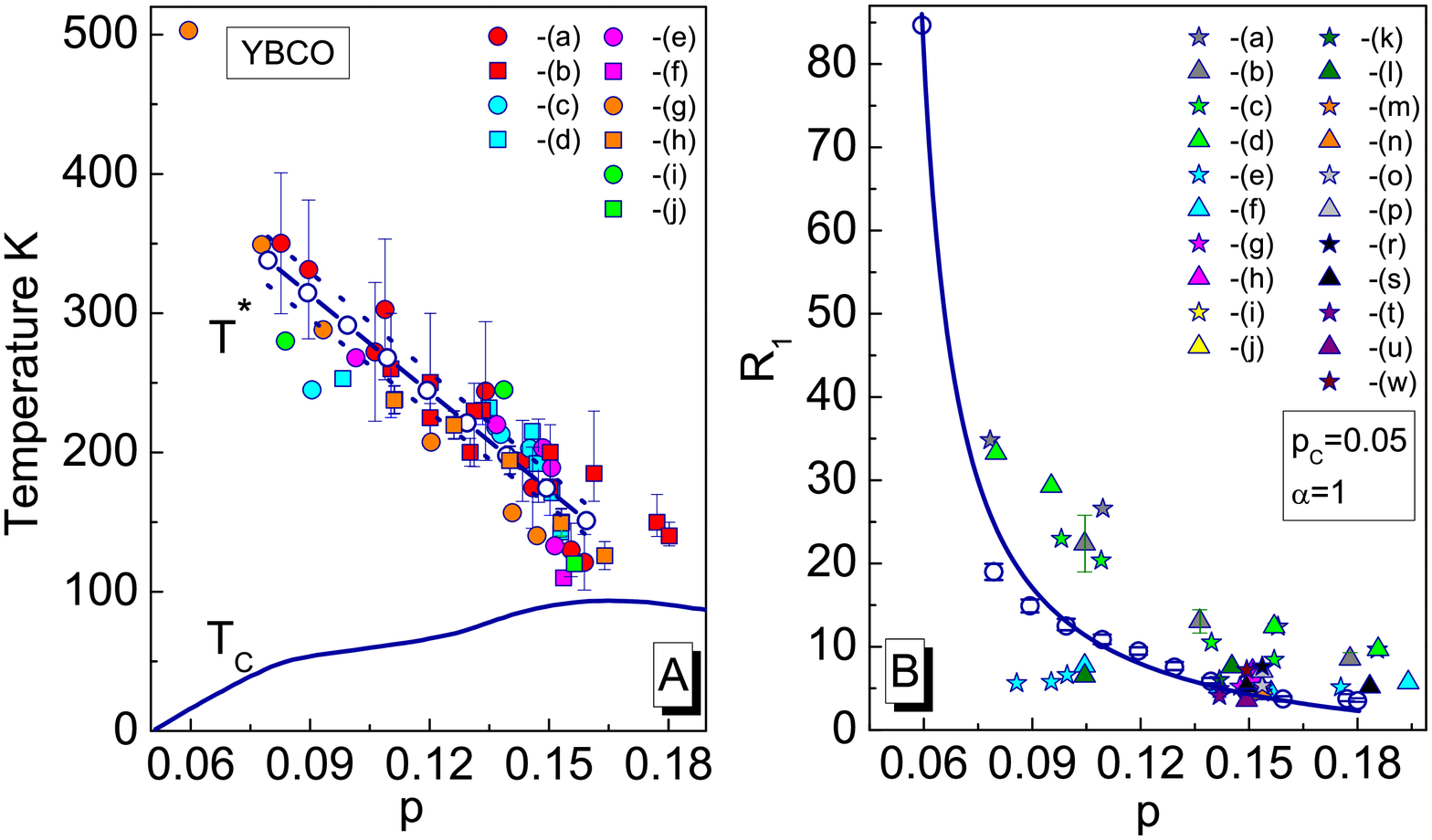}
\caption{(A) The temperatures $T_{C}$ and $T^{\star}$ as the functions of the doping in YBCO. The empty circles represent the averaged values of $T^{\star}$. The filled circles and squares are the experimental data:
(a)- Kabanov {\it et al.} \cite{Kabanov1999A}, 
(b)- Daou {\it et al.} \cite{Daou2010A}, 
(c)- Solovjov {\it et al.} \cite{Solovjov2009A}, 
(d)- Obolenskii {\it et al.} \cite{Obolenskii2006A}, 
(e)- Prokof'ev {\it et al.} \cite{Prokofev2003A}, 
(f)- Timusk {\it et al.} \cite{Timusk1999A}, 
(g)- Chaban \cite{Chaban2008A}, 
(h)- Vignolle {\it et al.} \cite{Vignolle2011A}, 
(i)- Lin {\it et al.} \cite{Lin2010A},
(j)- Rice {\it et al.} \cite{Rice1991A}.\\
(B) The dependence of the ratio $R_{1}$ on the doping for YBCO. The empty circles represent the numerical results. The line has been prepared on the basis of the formula \eq{r1}. The stars and the triangles are the experimental data: 
(a) - Sutherland {\it et al.} \cite{Sutherland2003A}, 
(b) - Nakayama {\it et al.} \cite{Nakayama2009A}, 
(c) - Kaminski {\it et al.} \cite{Kaminski2004A}, 
(d) - Plate {\it et al.} \cite{Plate2005A}, 
(e) - Morr {\it et al.} \cite{Morr1998A}, Fong {\it et al.} \cite{Fong1997A},
(f) - Yeh {\it et al.} \cite{Yeh2001A},
(g) - Born {\it et al.} \cite{Born2002A},
(h) - Murakami {\it et al.} \cite{Murakami2001A},
(i) - Edwards {\it et al.} \cite{Edwards1992A},
(j) - Edwards {\it et al.} \cite{Edwards1995A},
(k) - Tsai {\it et al.} \cite{Tsai1989A},
(l) - Schrieffer {\it et al.} \cite{Schrieffer2007A},
(m) - Maggio-Aprile {\it et al.} \cite{Maggio1995A}, \cite{Maggio1997A},
(n) - Koinuma {\it et al.} \cite{Koinuma1993A},
(o) - Koyanagi {\it et al.} \cite{Koyanagi1995A},
(p) - Kugler {\it et al.} \cite{Kugler2000A},
(r) - Nantoh {\it et al.} \cite{Nantoh1994A}, \cite{Nantoh1995A},
(s) - Shibata {\it et al.} \cite{Shibata2003A}, Ueno {\it et al.} \cite{Ueno2001A}, \cite{Ueno2003A},
(t) - Kirtley {\it et al.} \cite{Kirtley1987A},
(u) - Hoevers {\it et al.} \cite{Hoevers1988A},
(w) - Tanaka {\it et al.} \cite{Tanaka1994A}.
}
\label{f1}
\end{figure}
%

%
\begin{figure}[ht]
\includegraphics*[width=\columnwidth]{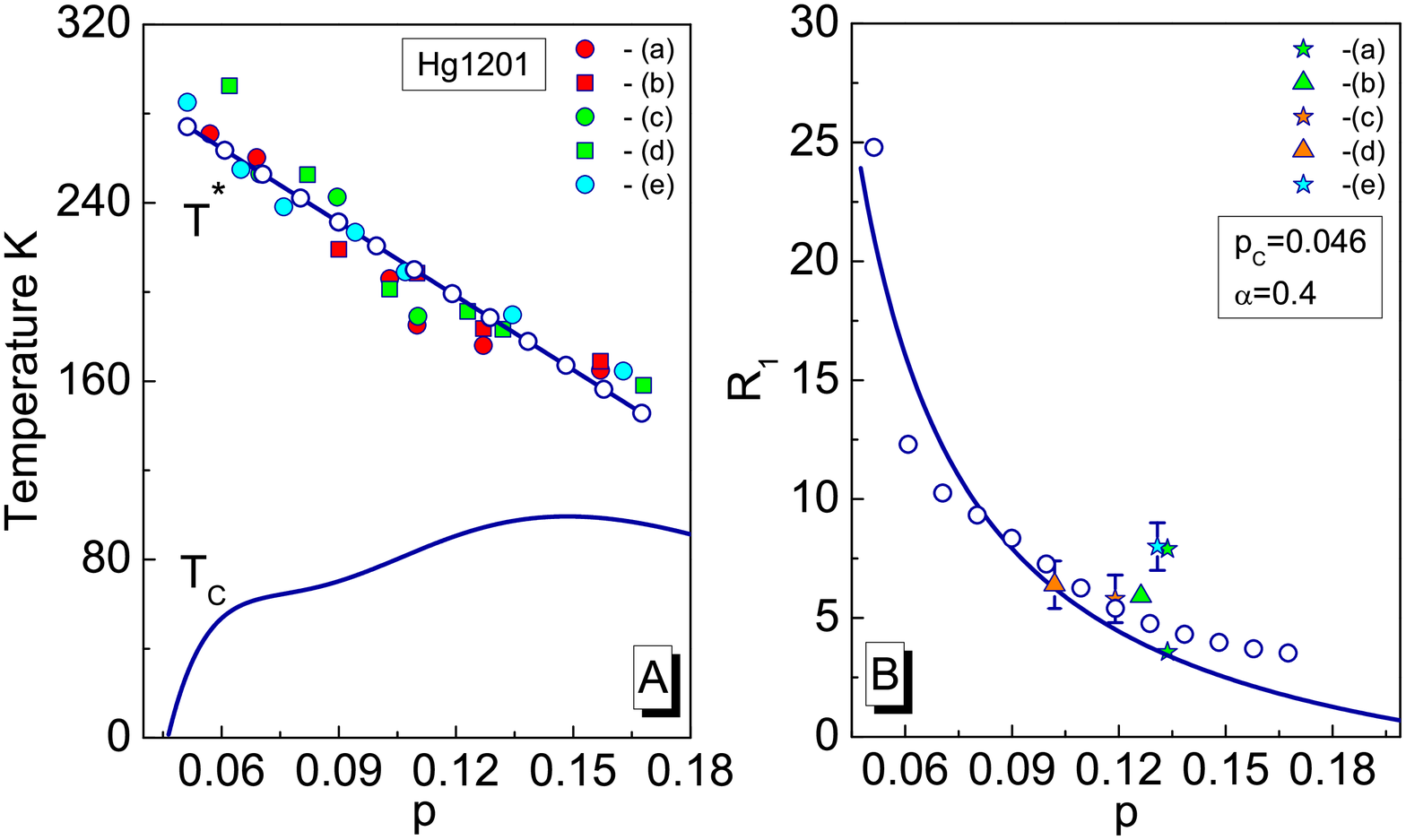}
\caption{(A) The temperatures $T_{C}$ and $T^{\star}$ as the functions of the doping in Hg1201. The empty circles represent the averaged values of $T^{\star}$. The filled circles and squares are the experimental data: 
(a), (b)- Yamamoto {\it et al.} \cite{Yamamoto2000A},
(c)- Bandyopadhyay {\it et al.} \cite{Bandyopadhyay2001A},
(d)- Yamamoto {\it et al.} \cite{Yamamoto2002A},
(e)- Honma  {\it et al.} \cite{Honma2004A}.\\
(B) The dependence of the ratio $R_{1}$ on the doping for Hg1201. The empty circles represent the strict numerical results. The line has been prepared on the basis of the formula \eq{r1}. The stars and the triangles are the experimental data: 
(a)- Wei {\it et al.} \cite{Wei1998A}, 
(b)- Hasegawa {\it et al.} \cite{Hasegawa1992A}, 
(c)- Yang {\it et al.} \cite{Yang2009A}, 
(d)- Guyard {\it et al.} \cite{Guyard2008A}, 
(e)- Inosov {\it et al.} \cite{Inosov2011A}. 
}
\label{f2}
\end{figure}
%

The first postulate emphasizes the importance of the quasi two-dimensionality of the system. The second one refers to the classical pairing mechanism given by Fr{\"o}hlich. The third postulate states that the strong electron correlations in the cuprates are inseparably coupled with the phonon subsystem. The first two postulates define the van Hove scenario \cite{Szczesniak2001A}, \cite{Szczesniak2006B}. The third postulate requires further discussion because it is far more subtle. In particular, it should be noted that the postulated electron correlations generalize the Hubbard approach; the classical two-body interaction is replaced with the three-body interaction (electron-electron-phonon). It should be pointed out very clearly that the third postulate does not require the additional phonon channel to be as strong as the electron channel.

By analyzing the set of the given postulates in the terms of the classical theory of the superconductivity \cite{Bardeen1957A}, \cite{Bardeen1957B}, it can be noticed that the first and second theory is based on the same philosophy. Namely, the physical system is constantly treated in a holistic manner, without the artificial division into the electron and phonon subsystem. 

The Hamiltonian corresponding to the presented postulates has been derived in the paper \cite{Szczesniak2012D}. Then, using the canonical transformation and the formalism of the thermodynamic Green functions one can get the equation determining the properties of the $d$-wave superconducting state. A detailed form of the equation is presented in the Appendix A.

The considered model has two main input parameters: the electron-phonon pairing potential $V^{\left(\eta\right)}$ and the electron-electron-phonon potential $U^{\left(\eta\right)}$. The first parameter can be calculated based on the value of the critical temperature ($T_{C}$), the second parameter has been chosen in such a way that it reproduces the value of the temperature at which the pseudogap appears ($T^{\star}$). The values of two other input parameters (the hopping integral and the characteristic phonon frequency) have been taken from the literature.

On the basis of the calculated input parameters, we have determined the ratio of the energy gap amplitude at the temperature of zero Kelvin to the critical temperature ($R_{1}\equiv 2\Delta_{tot}^{\left(0\right)}/k_{B}T_{C}$). The obtained results have been compared with the experimental data. In particular, the following compounds have been studied: ${\rm YBa_{2}Cu_{3}O_{7-y}}$ (YBCO), ${\rm HgBa_{2}CuO_{4+y}}$ (Hg1201), 
${\rm HgBa_{2}Cu_{1-x}Zn_{x}O_{4+y}}$ (Hg1201-Zn), and ${\rm HgBa_{2}Ca_{2}Cu_{3}O_{8+y}}$ (Hg1223). It should be noted that the considered compounds are the examples of the superconductors with the highest critical temperature (the optimal doping).

%
\begin{figure}[ht]
\includegraphics*[width=\columnwidth]{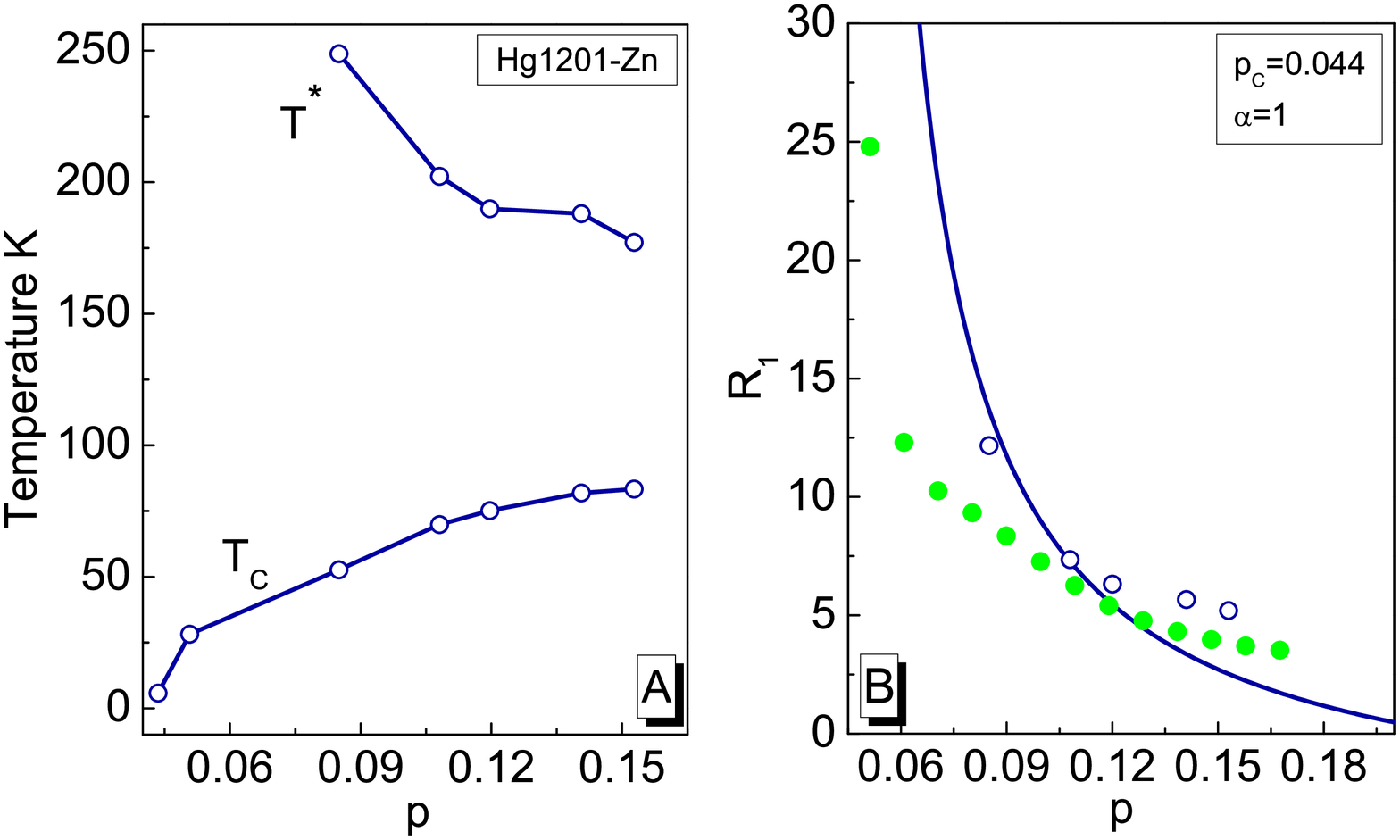}
\caption{(A) The temperatures $T_{C}$ and $T^{\star}$ as the functions of the doping in Hg1201-Zn. The diagram has been made on the basis of the paper by Yamamoto {\it et al.} \cite{Yamamoto2002A}.\\
(B) The dependence of the ratio $R_{1}$ on the doping. The empty circles represent the strict numerical results for Hg1201-Zn. The line has been prepared on the basis of the formula \eq{r1}. The green filled circles show the results achieved for Hg1201.
}
\label{f3}
\end{figure}
%

%
\begin{figure}[ht]
\includegraphics*[width=\columnwidth]{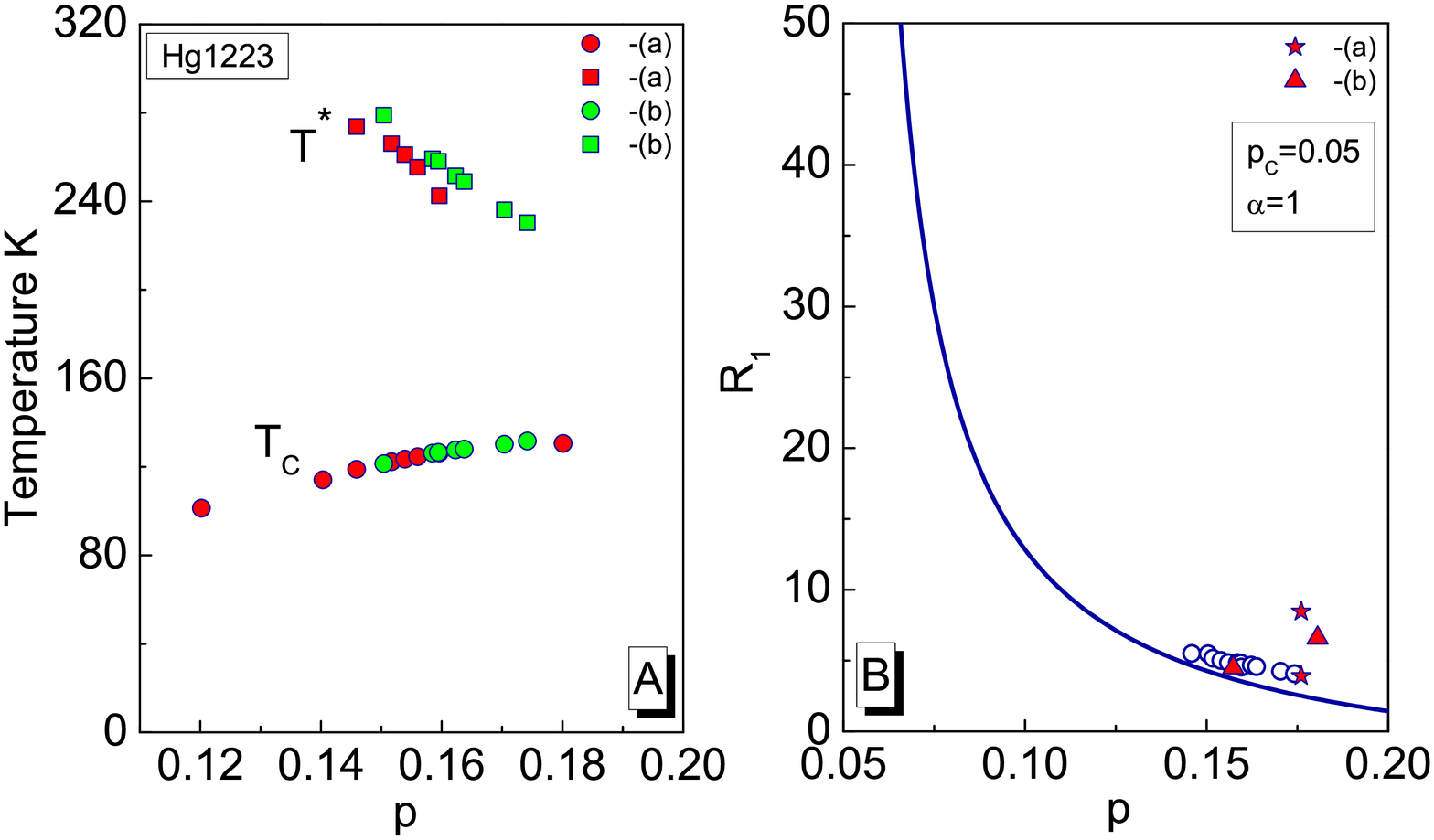}
\caption{(A) The temperatures $T_{C}$ and $T^{\star}$ as the functions of the doping in Hg1223. The diagram has been formed on the basis of the following experimental data: (a)- Liu {\it et al.} \cite{Liu2004A} and (b)- Lam {\it et al.} \cite{Lam2001A}.\\ 
(B) The dependence of the ratio $R_{1}$ on the doping for Hg1223. The empty circles represent the strict numerical results. The line has been prepared on the basis of the formula \eq{r1}. The experimental data: (a)- Jeong {\it et al.} \cite{Jeong1994A}, (b)- Rossel {\it et al.} \cite{Rossel1994A}.
}
\label{f4}
\end{figure}
%

In the first step, the ratio $R_{1}$ for the superconductor YBCO has been calculated. In \fig{f1} (A) there are presented the experimental courses of $T_{C}\left(p\right)$ and $T^{\star}\left(p\right)$, where $p$ denotes the concentration of the holes (the doping). The function $T_{C}\left(p\right)$ has been plotted on the basis of the data included in the paper \cite{Liang2006A}; the values $T^{\star}\left(p\right)$ have been determined averaging the large number of the experimental data. Then, on the basis of the equation (1A), the pairing potentials $V^{\left(\eta\right)}$ and $U^{\left(\eta\right)}$ have been calculated. The achieved results and the detailed descriptions of the experimental data used in this paper have been summarized in the Appendix B.

Next, we have prepared the dependence $R_{1}\left(p\right)$ (\fig{f1} (B)). It has been found that the theoretical results correctly reproduce the experimental data.

Let us notice that the dependence of the critical temperature on the doping for the cuprates has been parametrized with the help of the expression:  
$T_{C}\left(p\right)/T_{C,{\rm max}}=1-82.6\left(p-0.16\right)^{2}$ \cite{Presland1991A}. The shape of the function $R_{1}\left(p\right)$ can be reproduced in a similar way. In the considered case, the following result has been obtained:
\begin{equation}
\label{r1}
R_{1}\left(p\right)=\frac{p-5p_{c}}{\alpha p_{c}-p}\left[R_{1}\right]_{\rm dBCS},
\end{equation}
where $p_{c}$ is the value of the concentration of the holes at which the superconducting state appears; $\alpha$ is the fitting parameter. For YBCO, it has been assumed that: $p_{c}=0.05$ and $\alpha=1$. The symbol $\left[R_{1}\right]_{\rm dBCS}$ denotes the value of the ratio $R_{1}$ calculated  in the framework of the $d$-wave BCS model: $\left[R_{1}\right]_{\rm dBCS}=4.28$ \cite{Won1994A}. The form of the analytical results has been presented in \fig{f1} (B).

In the same way the course of the dependence of $R_{1}$ on $p$ in the family of the high temperature superconductors containing mercury has been submitted. The results have been presented in \fig{f2} - \fig{f4}; similarly for analysis as in the case of YBCO, in the Appendixes (C), (D) and (F), we have collected the values of the pairing potentials and the detailed descriptions of the experimental data.

In the case of the superconductor Hg1201, the shape of the function $T_{C}\left(p\right)$ has been obtained on the basis of the papers \cite{Yamamoto2000A}, \cite{Yamamoto2002A}, and \cite{Luo2008A}; the averaged values of $T^{\star}\left(p\right)$ have been determined with the help of the experimental data presented in \fig{f2} (A). The theoretical results and the experimental data have been collected in \fig{f2} (B). It has been found that the theoretical predictions agree with the experimental data. Then, by using the formula \eq{r1}, the course of the function $R_{1}\left(p\right)$ has been reproduced; the following parametrization has been selected: $p_{c}=0.046$ and $\alpha=0.4$. Also in this case the formula \eq{r1} correctly reproduces the dependence of the ratio $R_{1}$ on the concentration of the holes.

The superconductor Hg1201-Zn represents the case for which no experimental values of $R_{1}\left(p\right)$ have been found in the literature. However, there are well defined functions $T_{C}\left(p\right)$ and $T^{\star}\left(p\right)$ (\fig{f3} (A)). Thus, the course of the dependence $R_{1}\left(p\right)$ can be calculated theoretically. The strict numerical data have been collected in \fig{f3} (B). In addition, there have been presented the results obtained on the basis of the formula \eq{r1}, for which it has been assumed that: $p_{c}=0.044$ and $\alpha=1$. 

It is easy to notice that for $p\geq 0.11$, in the relation to the course obtained for Hg1201, the disorder induced by zinc does not exert the significant influence on $R_{1}\left(p\right)$. Much bigger deviations can be noticed for the lower values of the doping. The discussed effect can be relatively easily verified by the experimenters - we strongly encourage to do so.

The superconductor Hg1223 is characterized with the highest observed critical temperature for the optimal doping ($T_{C}\simeq 135$ K) \cite{Chu1993A}. The full courses of $T_{C}\left(p\right)$ and $T^{\star}\left(p\right)$ for Hg1223 have been determined in the paper \cite{Liu2004A} and \cite{Lam2001A} (\fig{f4} (A)). Using the already discussed method of the analysis, we have obtained the theoretical form of the function $R_{1}\left(p\right)$. The results have been presented in \fig{f4} (B). We have found that the theoretical predictions reproduce the existing experimental data very well. Additionally, in \fig{f4} (B) there have been plotted the analytical values of $R_{1}\left(p\right)$; we have assumed that: $p_{c}=0.05$ and $\alpha=1$.

To summarize: in the presented paper, we have calculated the dependence of the ratio $R_{1}$ on the doping for four superconductors characterized by high values of the critical temperature. In the case of YBCO, Hg1201 and Hg1223, the theoretical results reproduce the experimental data very well. For the superconductor Hg1201-Zn only the theoretical results have been provided due to the fact that the relevant experiments have not been undertaken yet. Finally, we have shown that the shape of the function $R_{1}\left(p\right)$ can be reproduced by the analytical formula. 

The presented theory can be used to explore all other thermodynamic parameters of the cuprates. It seems that one of the most fundamental issues is associated with the interpretation of the experimental data obtained by using the ARPES method. Such studies are currently underway and, in a short time, the results will be presented.

\begin{acknowledgments}
The authors wish to thank Prof. K. Dzili{\'n}ski, the Head of the Institute of Physics at Cz{\c{e}}stochowa University of Technology, for providing excellent working conditions and the financial support. Additionally, we would like to thank our colleagues: D. Szcz{\c{e}}{\'s}niak, M.W. Jarosik, E.A. Drzazga, and T. Mila for their kindness and the technical support offered during the preparation of this paper.

The calculations have been conducted on the Cz{\c{e}}stochowa University of Technology cluster, built in the framework of the
PLATON project, no. POIG.02.03.00-00-028/08 - the service of the campus calculations U3.
\end{acknowledgments}

\appendix

\begin{widetext}

\section{THE FUNDAMENTAL EQUATION}

The equation for the high temperature superconducting state has the form \cite{Szczesniak2012D} 
(see also \cite{Szczesniak2012E} and \cite{Szczesniak2012F}):
\begin{equation}
\label{rA1}
1=\left(V^{\left(\eta\right)}+\frac{U^{\left(\eta\right)}}{6}|\Delta^{\left(\eta\right)}|^{2}\right)\frac{1}{N_{0}}\sum^{\omega_{0}}_{\kvec}
\frac{\eta^{2}(\bf k)}{2E_{\kvec}^{\left(\eta\right)}}\tanh\frac{\beta E_{\kvec}^{\left(\eta\right)}}{2},
\end{equation}
where the quantities $V^{\left(\eta\right)}$ and $U^{\left(\eta\right)}$ denote the effective pairing potentials for the electron-phonon and 
electron-electron-phonon interaction, respectively. The symbol $\Delta^{\left(\eta\right)}$ is the amplitude of the $d$-wave order parameter; $\eta{\left(\bf k\right)}\equiv 2\left[\cos\left(k_{x}\right)-\cos\left(k_{y}\right)\right]$. The function $E_{\kvec}^{\left(\eta\right)}$ is given by the expression:
\begin{equation}
\label{rA2}
E_{\kvec}^{\left(\eta\right)}\equiv\sqrt{\varepsilon_{\kvec}^{2}+\left(V^{\left(\eta\right)}+ \frac{U^{\left(\eta\right)}}{6}|\Delta^{\left(\eta\right)}|^{2}\right)^{2}\left(|\Delta^{\left(\eta\right)}|\eta\left(\bf k\right)\right)^{2}}, 
\end{equation}
where $\varepsilon_{\kvec}$ denotes the electron band energy: $\varepsilon_{\kvec}=-t\gamma\left(\bf k\right)$; $t$ is the hopping integral and $\gamma\left({\bf k}\right)\equiv 2\left [\cos\left(k_{x}\right)+\cos\left(k_{y}\right)\right]$. The quantity $\beta$ has been defined as: $\beta\equiv 1/k_{B}T$, where $k_{B}$ is the Boltzmann constant.

The equation (\ref{rA1}) has too much a complicated form to be solved in the analytical way. For this reason, it has been analyzed by using the numerical methods. Additionally, we have assumed that: $\sum^{\omega_{0}}_{\kvec}\simeq \int^{\pi}_{-\pi}\int^{\pi}_{-\pi}dk_{x}dk_{y}\theta\left(\omega_{0}-\left|\varepsilon_{\left(\kvec_{x},\kvec_{y}\right)}\right|\right)$, where $\theta$ is the Heaviside function. The normalization constant is given by the following: $N_{0}\equiv 1/\sum^{\omega_{0}}_{\kvec}$; the symbol $\omega_{0}$ represents the characteristic phonon frequency, which is of the order of Debye frequency.

The energy gap amplitude at the temperature of zero Kelvin is defined as: $\Delta^{\left(0\right)}_{tot}\equiv\left(V^{\left(\eta\right)}+\frac{U^{\left(\eta\right)}}{6}
\left|\Delta^{\left(\eta\right)}_{0}\right|^{2}\right)\left|\Delta^{\left(\eta\right)}_{0}\right|$.
\end{widetext}

\clearpage
\begin{widetext}
\section{THE VALUES OF THE PAIRING POTENTIALS $V^{\left(\eta\right)}$, $U^{\left(\eta\right)}$ AND THE EXPERIMENTAL RESULTS FOR YBCO}

%
\begin{table*}[ht]
\caption{\label{t1} The values of the potentials $V^{\left(\eta\right)}$ and $U^{\left(\eta\right)}$ calculated on the basis of $T_{C}$ and $T^{\star}$.}
\begin{ruledtabular}
\begin{tabular}{ccccccccccc}
Material & Hole doping &$t$ (meV)&Ref.&$\omega_{0}$ (meV)&Ref.&$T_{C}$ (K)&$T^{\star}$ (K)&$\left(T^{\star}-T_{C}\right)/T_{C}$&$V^{\left(\eta\right)}$ (meV)&$U^{\left(\eta\right)}$ (meV)\\
         &      &         &    &      &    &    &           &    &                             &                             \\
\hline
YBCO	&$p=	0.059	$\footnote{The points outside the main line.}&	$250$&	\cite{Xu1987A}&	$75$&	\cite{Kim1998A}&	$	15.48	$&$	502.99	$&$	31.49	$&$	{\bf	2.677}	$&$	{\bf	115.897}$\\
	&$p=	0.079	$&	&	&	&	&	$	45.40	$&$	338.12	$&$	6.45	$&$	{\bf	4.085}	$&$	{\bf	76.106}$\\
	&$p=	0.089	$&	&	&	&	&	$	53.60	$&$	314.74	$&$	4.87	$&$	{\bf	4.394}	$&$	{\bf	70.376}$\\
	&$p=	0.099	$&	&	&	&	&	$	57.89	$&$	291.35	$&$	4.03	$&$	{\bf	4.543}	$&$	{\bf	64.825}$\\
	&$p=	0.109	$&	&	&	&	&	$	61.93	$&$	267.96	$&$	3.33	$&$	{\bf	4.709}	$&$	{\bf	59.773}$\\
	&$p=	0.119	$&	&	&	&	&	$	66.48	$&$	244.58	$&$	2.68	$&$	{\bf	4.887}	$&$	{\bf	55.934}$\\
	&$p=	0.129	$&	&	&	&	&	$	73.10	$&$	221.19	$&$	2.03	$&$	{\bf	5.109}	$&$	{\bf	50.057}$\\
	&$p=	0.139	$&	&	&	&	&	$	82.75	$&$	197.81	$&$	1.39	$&$	{\bf	5.437}	$&$	{\bf	43.195}$\\
	&$p=	0.149	$&	&	&	&	&	$	90.02	$&$	174.42	$&$	0.94	$&$	{\bf	5.679}	$&$	{\bf	37.027}$\\
	&$p=	0.159	$&	&	&	&	&	$	93.78	$&$	151.03	$&$	0.61	$&$	{\bf	5.797}	$&$	{\bf	31.747}$\\
	&$p=	0.177^a$&	&	&	&	&	$	92.32	$&$	149.85	$&$	0.62	$&$	{\bf	5.752}	$&$	{\bf	31.762}$\\
	&$p=	0.180^a$&	&	&	&	&	$	91.21	$&$	140.11	$&$	0.54	$&$	{\bf	5.717}	$&$	{\bf	29.910}$\\
\end{tabular}
\end{ruledtabular}
\end{table*}
%

%
\begin{table*}[ht]
\caption{\label{t2} The experimental values of $T_{C}$ and $T^{\star}$ for YBCO.}
\begin{ruledtabular}
\begin{tabular}{cccc}
Hole doping&$T_{C}$ (K)&$T^{\star}$ (K)&Ref.\\
\hline
$p=0.159	$	&	93.69	&	$121.12\pm 19.92$	&	\cite{Kabanov1999A}\\
$p=0.156	$	&	92.69	&	$130\pm	19.16$&	\\
$p=0.146	$	&	88.92	&	$174.68\pm	29.89$	&	\\
$p=0.143	$	&	87.96	&	$193.92\pm	29.12$	&	\\
$p=0.147	$	&	86.1	&	$194.1\pm	29.12$	&	\\
$p=0.134	$	&	77.68	&	$244.08\pm	49.81$	&	\\
$p=0.106	$	&	61.69	&	$272.19\pm	50.57$	&	\\
$p=0.109	$	&	60.7	&	$302.74\pm	49.81$	&	\\
$p=0.090	$	&	53.57	&	$331.33\pm	49.81$	&	\\
$p=0.083	$	&	49	&	   $350.2\pm	50.57$	&	\\
\hline
$p=	0.153 $&	92.57 & $149.95\pm 10$	& 	\cite{Daou2010A}\\
$p=	0.140	$&	84.33 & $194.73\pm 10$	& \\
$p=	0.130	$&	73.67 & $200.12\pm 10$	& \\
$p=	0.126	$&	70.27 & $220.07\pm 10$	& \\
$p=	0.111 $&	62.52 & $238.12\pm 10$	& \\\\

$p=	0.150 $&	91.11	&	$175.24 (+20;-19)$&	\\
$p=	0.133	$&	76.58	&	$230.25 (+19;-10;)$&	\\
$p=	0.120	$&	66.40	&	$250.24 (+50;+45)$&  \\
$p=	0.110 $&	62.04	&	$260.01 (+40;-35)$&	\\\\

$p=	0.180 $&	91.21	&	$140.11	(+10;-5)$&	\\
$p=	0.177	$&	92.32	&	$149.85	(+20;-10)$&	\\
$p=	0.161	$&	94.00	&	$184.92	(+45;-20)$&	\\
$p=	0.150	$&	90.59	&	$200.04	(+20;-20)$&  \\
$p=	0.131	$&	74.76	&	$229.78	(+20;-40)$&  \\
$p=	0.120 $&	66.96	&	$224.96	(+10;-20)$&	\\
\hline
$p=0.145$    & 87.4  & 203  &  \cite{Solovjov2009A}\\
$p=0.138$    & 81.4  & 213  &                 \\
$p=0.137$    & 80.3  & 218  &                 \\
$p=0.090$    & 54.2  & 245  &                 \\
\hline
$p=0.153$   & 91.74  & 143  &  \cite{Obolenskii2006A}\\
$p=0.151$   & 90.85  & 171  &                 \\
$p=0.147$   & 88.71  & 192  &                 \\
$p=0.146$   & 87.89  & 215  &                 \\
$p=0.135$   & 78.52  & 232  &                 \\
$p=0.098$   & 57.45  & 253  &                 \\
\hline
$p=0.151$   & 91.2   & 133   &  \cite{Prokofev2003A}\\
$p=0.150$   & 90.8   & 189   &                  \\
$p=0.148$   & 89.5   & 203.3 &                  \\
$p=0.137$   & 80.5   & 220   &                  \\
$p=0.101$   & 58.7   & 268   &                  \\
\hline
$p=0.154$	&	92	&	110	&	\cite{Timusk1999A}\\
\hline
$p=0.147$    & 88.71  & 140.30  &  \cite{Chaban2008A}\\
$p=0.141$    & 84.19  & 156.72  &                 \\
$p=0.120$    & 67.10  & 207.46  &                 \\
$p=0.093$    & 55.48  & 288.06  &                 \\
$p=0.078$    & 43.23  & 349.25  &                 \\
$p=0.059$    & 15.48  & 502.99  &                 \\
\hline
$p=	0.164	$&	91.92	&$	125.92	\pm	10	$&  \cite{Vignolle2011A}\\
$p=	0.153	$&	89.94	&$	149.23	\pm	10	$&\\
$p=	0.140	$&	82.17	&$	194.11	\pm	10	$&\\
$p=	0.126	$&	69.02	&$	219.52	\pm	10	$&\\
$p=	0.111	$&	61.27	&$	237.47	\pm	10	$&\\
\hline
$p=0.138$    & 81.76  & 245  &  \cite{Lin2010A}\\
$p=0.084$    & 50     & 280  &  \\
\hline
$p=0.156$    & 93     & 120  &  \cite{Rice1991A}\\
\end{tabular}
\end{ruledtabular}
\end{table*}
%

%
\begin{table*}[ht]
\caption{\label{t3} The experimental values of the energy gap amplitude and the $R_{1}$ parameter for YBCO.}
\begin{ruledtabular}
\begin{tabular}{ccccc}
Hole doping &$T_{C}$ (K)&$\Delta_{tot}^{\left(0\right)}$ (meV)&$R_{1}$&Ref.\\
\hline
$p=0.078$ & 44   & 66 & 34.81 & \cite{Sutherland2003A}\\
$p=0.110$ & 62   & 71 & 26.58 &                  \\
$p=0.158$ & 93.5 & 50 & 12.41 &                  \\
$p=0.186$ & 89   & 37 & 9.65  &                  \\
\hline
$p=0.105$ & 60   & $58\pm 8.8$  & $22.37\pm 3.39$ &  \cite{Nakayama2009A}\\
$p=0.136$ & 80   & $45\pm 4.9$  & $13.04\pm 1.40$ &                 \\
$p=0.178$ & 92   & $34\pm 3.1$  & $8.55 \pm 0.77$ &                 \\
\hline
$p=0.098$ & 57.4 & 56.8 & 22.96 & \cite{Kaminski2004A}\\
$p=0.109$ & 61.8 & 54.2 & 20.36 &                \\
$p=0.140$ & 83   & 37.6 & 10.52 &                \\
$p=0.186$ & 93.2 & 33.8 & 8.42  &                \\
\hline
$p=0.080$  & 46.2 & 66.3   & 33.27 & \cite{Plate2005A}\\
$p=0.095$  & 56.3 & 71     & 29.3  &             \\
$p=0.157$  & 93.2 & 49.9   & 12.42 &             \\
$p=0.186$  & 88.9 & 37.2   & 9.7   &             \\
\hline
$p=0.086$  &  51.6  &  12.5  &  5.62  & \cite{Morr1998A}, \cite{Fong1997A}\\
$p=0.095$  &  56.3  &  14    &  5.77  &                   \\
$p=0.100$  &  58    &  16.5  &  6.6   &                   \\
$p=0.175$  &  93    &  20.5  &  5.12  &                   \\
\hline
$p=0.105$  &  60    &  $\sim 20$  & 7.7 & \cite{Yeh2001A}\\
$p=0.156$  &  92.9  &  18         & 4.5 &           \\
$p=0.142$  &   85  &   19        & 5.2 &           \\
$p=0.194$  &  78  &    19      & 5.7 &           \\

\hline
$p=0.147$  &  89    &  20         & 5.2 & \cite{Born2002A}\\
\hline
$p=0.151$  &  91  &  24-32                                                              & 7.1 & \cite{Murakami2001A}\\
$p=0.151$  &  91  &  $\sim 25$                                                          & 6.4 &                \\
\hline
$p=0.154$  &  92  &  $30\pm 8$  & 7.6   & \cite{Edwards1992A}\\
\hline
$p=0.154$  &  92  &  20         & 5     & \cite{Edwards1995A}\\  
\hline
$p=0.142$  &  85  &  22         & 6     & \cite{Tsai1989A}\\  
\hline
$p=0.105$  &  60    &  17       & 6.5      & \cite{Schrieffer2007A}\\  
$p=0.145$  &  87.5  &  29       & 7.6      &\\
\hline
$p=0.154$  &  60    &  $20\pm 2$       & 5      & \cite{Maggio1995A},\cite{Maggio1997A}\\  
$p=0.154$  &  87.5  &  18       & 4.5      &\\
\hline
$p=0.154$  &  92    &  17.5       & 4.4      & \cite{Koinuma1993A}\\
\hline
$p=0.154$  &  92    &  $\sim 21$       & 5.3      & \cite{Koyanagi1995A}\\
\hline
$p=0.154$  &  92    &  $28\pm 2$       & 7.1      & \cite{Kugler2000A}\\
\hline
$p=0.154$  &  92    &  $30\pm 10$       & 7.6      & \cite{Nantoh1994A}, \cite{Nantoh1995A}\\ 
\hline
$p=0.183$  &  90    &  20       & 5.2      & \cite{Shibata2003A}\\ 
\hline
$p=0.149$  &  90    &   $\sim 20$        & 5.2      & \cite{Ueno2001A},\cite{Ueno2003A}\\ 
\hline
$p=0.142$  &  85    &  18       & 4.9      & \cite{Kirtley1987A}\\  
$p=0.142$  &  85    &  15       & 4.1      &\\
\hline
$p=0.149$  &  90    &  $14\pm 2$       & 3.6      & \cite{Hoevers1988A}\\  
\hline
$p=0.149$  &  90    &  28       & 7.2      & \cite{Tanaka1994A}\\  
\end{tabular}
\end{ruledtabular}
\end{table*}
%
\end{widetext}
\clearpage
\begin{widetext}
\section{The values of the potentials $V^{\left(\eta\right)}$ and $U^{\left(\eta\right)}$ calculated on the basis of $T_{C}$ and $T^{\star}$ for Hg1201}

%
\begin{table*}[ht]
\caption{\label{t4} The values of the potentials $V^{\left(\eta\right)}$ and $U^{\left(\eta\right)}$ calculated on the basis of  $T_{C}$ and $T^{\star}$.}
\begin{ruledtabular}
\begin{tabular}{ccccccccccc}
Material & Hole doping &$t$ (meV)&Ref.&$\omega_{0}$ (meV)&Ref.&$T_{C}$ (K)&$T^{\star}$ (K)&$\left(T^{\star}-T_{C}\right)/T_{C}$&$V^{\left(\eta\right)}$ (meV)&$U^{\left(\eta\right)}$ (meV)\\
         &      &         &    &      &    &    &           &    &                             &                             \\
\hline
Hg1201	&$p=	0.051	$&	$450$&	\cite{Pavarini2001A}&	$95$&	\cite{Dovhyj2011A}&	$	28.93	$&$	274.21	$&$	8.48	$&$	{\bf	3.835}	$&$	{\bf	66.551}$\\
&$p=	0.061	$&	&	&	&	&	$	54.84	$&$	263.60	$&$	3.81	$&$	{\bf	4.965	}$&$	{\bf	59.881	}$\\
&$p=	0.071	$&	&	&	&	&	$	62.67	$&$	252.88	$&$	3.04	$&$	{\bf	5.260	}$&$	{\bf	56.754	}$\\
&$p=	0.080	$&	&	&	&	&	$	65.97	$&$	242.16	$&$	2.67	$&$	{\bf	5.377	}$&$	{\bf	54.577	}$\\
&$p=	0.090	$&	&	&	&	&	$	70.21	$&$	231.44	$&$	2.30	$&$	{\bf	5.503	}$&$	{\bf	52.168	}$\\
&$p=	0.100	$&	&	&	&	&	$	76.34	$&$	220.72	$&$	1.89	$&$	{\bf	5.755	}$&$	{\bf	49.212	}$\\
&$p=	0.109	$&	&	&	&	&	$	83.40	$&$	210.00	$&$	1.52	$&$	{\bf	5.989	}$&$	{\bf	46.309	}$\\
&$p=	0.119	$&	&	&	&	&	$	90.01	$&$	199.28	$&$	1.21	$&$	{\bf	6.202	}$&$	{\bf	43.326	}$\\
&$p=	0.129	$&	&	&	&	&	$	95.14	$&$	188.56	$&$	0.98	$&$	{\bf	6.371	}$&$	{\bf	40.621	}$\\
&$p=	0.138	$&	&	&	&	&	$	98.26	$&$	177.84	$&$	0.81	$&$	{\bf	6.460	}$&$	{\bf	38.385	}$\\
&$p=	0.148	$&	&	&	&	&	$	99.32	$&$	167.12	$&$	0.68	$&$	{\bf	6.491	}$&$	{\bf	36.275	}$\\
&$p=	0.158	$&	&	&	&	&	$	98.51	$&$	156.40	$&$	0.59	$&$	{\bf	6.468	}$&$	{\bf	34.517	}$\\
&$p=	0.168	$&	&	&	&	&	$	96.13	$&$	145.68	$&$	0.52	$&$	{\bf	6.400	}$&$	{\bf	33.034	}$\\
\end{tabular}
\end{ruledtabular}
\end{table*}
%

%
\begin{table*}[ht]
\caption{\label{t5} The experimental values of $T_{C}$ and $T^{\star}$ for Hg1201.}
\begin{ruledtabular}
\begin{tabular}{cccc}
Hole doping&$T_{C}$ (K)&$T^{\star}$ (K)&Ref.\\
\hline
$p=0.057 $& 46 &  270.93   &  \cite{Yamamoto2000A}\\
$p=0.069	$&	62	&	260.28	&	\\
$p=0.090	$&	72	&	231.35	&	\\
$p=0.103	$&	77	&	205.97	&	\\
$p=0.110	$&	83	&	185.15	&	\\
$p=0.127	$&	95	&	176.03	&	\\
$p=0.157	$&	98	&	164.90	&	\\\\
$p=0.090	$&	72	&	219.16	&	\\
$p=0.110	$&	83	&	208.52	&	\\
$p=0.127	$&	95	&	183.65	&	\\
$p=0.157	$&	98	&	168.97	&	\\
\hline
$p=	0.070	$&	62	&	252.98	&	\cite{Bandyopadhyay2001A}\\
$p=	0.090	$&	70	&	242.58	&	\\
$p=	0.110	$&	83	&	189.13	&	\\
\hline
$p=	0.062	$&	48	&	292.54	&	\cite{Yamamoto2002A}\\
$p=	0.082	$&	72	&	252.63	&	\\
$p=	0.103	$&	82	&	201.29	&	\\
$p=	0.123	$&	89	&	191.35	&	\\
$p=	0.132	$&	95	&	183.28	&	\\
$p=	0.168	$&	98	&	158.14	&	\\
\hline
$p=	0.051	$&	29	&	285.17	&	\cite{Honma2004A}\\
$p=	0.065	$&	59	&	255.02	&	\\
$p=	0.076	$&	65	&	238.22	&	\\
$p=	0.094	$&	73	&	226.85	&	\\
$p=	0.107	$&	82	&	209.06	&	\\
$p=	0.134	$&	97	&	189.79	&	\\
$p=	0.163	$&	97	&	164.58	&	\\
\end{tabular}
\end{ruledtabular}
\end{table*}
%

%
\begin{table*}[ht]
\caption{\label{t6} The experimental values of the energy gap amplitude and the $R_{1}$ parameter for Hg1201.}
\begin{ruledtabular}
\begin{tabular}{ccccc}
Hole doping &$T_{C}$ (K)&$\Delta_{tot}^{\left(0\right)}$ (meV)&$R_{1}$&Ref.\\
\hline
$p=0.134$ & 97   & 14.92 & 3.57 & \cite{Wei1998A}\\
$p=0.134$ & 97   & 33.02 & 7.90 &                  \\
\hline
$p=0.126$ & 94 & 23.98 & 5.92 & \cite{Hasegawa1992A}\\
\hline
$p=0.119$  & 90 & 22.49   & $5.80\pm 1$ & \cite{Yang2009A}\\
\hline
$p=0.102$  &  78  &  21.5  &  $6.4\pm 1$  & \cite{Guyard2008A}\\
\hline
$p=0.131$  &  96    &  33.09  & $8\pm 1$ & \cite{Inosov2011A}\\
\end{tabular}
\end{ruledtabular}
\end{table*}
%
\end{widetext}

\clearpage
\begin{widetext}
\section{THE VALUES OF THE PAIRING POTENTIALS $V^{\left(\eta\right)}$, $U^{\left(\eta\right)}$ AND THE EXPERIMENTAL RESULTS FOR Hg1201-Zn}

%
\begin{table*}[ht]
\caption{\label{t7} The values of the potentials $V^{\left(\eta\right)}$ and $U^{\left(\eta\right)}$ calculated on the basis of $T_{C}$ and $T^{\star}$.}
\begin{ruledtabular}
\begin{tabular}{ccccccccccc}
Material & Hole doping &$t$ (meV)&Ref.&$\omega_{0}$ (meV)&Ref.&$T_{C}$ (K)&$T^{\star}$ (K)&$\left(T^{\star}-T_{C}\right)/T_{C}$&$V^{\left(\eta\right)}$ (meV)&$U^{\left(\eta\right)}$ (meV)\\
         &      &         &    &      &    &    &           &    &                             &                             \\
\hline
Hg1201-Zn	&$p=	0.085	$&	$450$&	\cite{Pavarini2001A}&	$95$&	\cite{Dovhyj2011A}&	$	52.65	$&$	248.84	$&$	3.73	$&$	{\bf	4.848}	$&$	{\bf	58.090}$\\
&$p=	0.108	$&	&	&	&	&	$	69.87	$&$	202.25	$&$	1.89	$&$	{\bf	5.488	}$&$	{\bf	47.766	}$\\
&$p=	0.120	$&	&	&	&	&	$	75.14	$&$	189.91	$&$	1.53	$&$	{\bf	5.710	}$&$	{\bf	44.605	}$\\
&$p=	0.141	$&	&	&	&	&	$	81.88	$&$	188.07	$&$	1.30	$&$	{\bf	5.946	}$&$	{\bf	43.026	}$\\
&$p=	0.153	$&	&	&	&	&	$	83.34	$&$	177.15	$&$	1.13	$&$	{\bf	5.987	}$&$	{\bf	41.045	}$\\
\end{tabular}
\end{ruledtabular}
\end{table*}
%

%
\begin{table*}[ht]
\caption{\label{t8} The experimental values of $T_{C}$ and $T^{\star}$ for Hg1201-Zn.}
\begin{ruledtabular}
\begin{tabular}{ccccc}
Hole doping&$T_{C}$ (K)&$T^{\star}$ (K)&Ref.\\
\hline
$p=	0.085	$&	52.65	   &	248.84	&  \cite{Yamamoto2002A}\\
$p=	0.108	$&	69.87	   &	202.25	&	\\
$p=	0.120	$&	75.14	&	189.91	&	\\
$p=	0.141	$&	81.88	&	188.07	&	\\
$p=	0.153	$&	83.34	&	177.15	&	\\
\end{tabular}
\end{ruledtabular}
\end{table*}
%
\end{widetext}
\clearpage
\begin{widetext}
\section{THE VALUES OF THE PAIRING POTENTIALS $V^{\left(\eta\right)}$, $U^{\left(\eta\right)}$ AND THE EXPERIMENTAL RESULTS FOR Hg1223}

%
\begin{table*}[ht]
\caption{\label{t9} The values of the potentials $V^{\left(\eta\right)}$ and $U^{\left(\eta\right)}$ calculated on the basis of $T_{C}$ and $T^{\star}$.}
\begin{ruledtabular}
\begin{tabular}{ccccccccccc}
Material & Hole doping &$t$ (meV)&Ref.&$\omega_{0}$ (meV)&Ref.&$T_{C}$ (K)&$T^{\star}$ (K)&$\left(T^{\star}-T_{C}\right)/T_{C}$&$V^{\left(\eta\right)}$ (meV)&$U^{\left(\eta\right)}$ (meV)\\
         &      &         &    &      &    &    &           &    &                             &                             \\
\hline
Hg1223	&$p=	0.1459	$&	$450$&	\cite{Pavarini2001A}&	$95$&	\cite{Dovhyj2011A}&	$	118.90	$&$	273.80	$&$	1.30	$&$	{\bf	7.157}	$&$	{\bf	50.891}$\\
&$p=	0.1504	$&	&	&	&	&	$	121.46	$&$	278.90	$&$	1.30	$&$	{\bf	7.241	}$&$	{\bf	51.368	}$\\
&$p=	0.1517	$&	&	&	&	&	$	122.30	$&$	266.10	$&$	1.18	$&$	{\bf	7.264	}$&$	{\bf	48.996	}$\\
&$p=	0.1539	$&	&	&	&	&	$	123.50	$&$	261.10	$&$	1.11	$&$	{\bf	7.297	}$&$	{\bf	47.925	}$\\
&$p=	0.1560	$&	&	&	&	&	$	124.60	$&$	255.40	$&$	1.05	$&$	{\bf	7.326	}$&$	{\bf	46.760	}$\\
&$p=	0.1585	$&	&	&	&	&	$	126.21	$&$	259.18	$&$	1.05	$&$	{\bf	7.370	}$&$	{\bf	47.205	}$\\
&$p=	0.1594	$&	&	&	&	&	$	126.58	$&$	258.08	$&$	1.04	$&$	{\bf	7.381	}$&$	{\bf	46.971	}$\\
&$p=	0.1596	$&	&	&	&	&	$	126.30	$&$	242.50	$&$	0.92	$&$	{\bf	7.372	}$&$	{\bf	44.411	}$\\
&$p=	0.1623	$&	&	&	&	&	$	127.67	$&$	251.51	$&$	0.97	$&$	{\bf	7.414	}$&$	{\bf	45.994	}$\\
&$p=	0.1637	$&	&	&	&	&	$	128.04	$&$	248.95	$&$	0.94	$&$	{\bf	7.426	}$&$	{\bf	45.104	}$\\
&$p=	0.1704	$&	&	&	&	&	$	130.23	$&$	236.16	$&$	0.81	$&$	{\bf	7.491	}$&$	{\bf	42.573	}$\\
&$p=	0.1742	$&	&	&	&	&	$	131.69	$&$	230.32	$&$	0.75	$&$	{\bf	7.534	}$&$	{\bf	41.387	}$\\
\end{tabular}
\end{ruledtabular}
\end{table*}
%

%
\begin{table*}[ht]
\caption{\label{t10} The experimental values of $T_{C}$ and $T^{\star}$ for Hg1223.}
\begin{ruledtabular}
\begin{tabular}{ccccc}
Hole doping&$T_{C}$ (K)&$T^{\star}$ (K)&Ref.\\
\hline
$p=	0.146	$&118.9	& 273.8&  \cite{Liu2004A}\\
$p=	0.152	$&122.3	& 266.1	&	\\
$p=	0.154	$&123.5	& 261.1	&	\\
$p=	0.156	$&124.6	& 255.4	&	\\
$p=	0.160	$&126.3	& 242.5	&	\\
\hline
$p=	0.150	$&120.9 &	279.0&  \cite{Lam2001A}\\
$p=	0.159	$&125.9 &	259.0	&	\\
$p=	0.163	$&127.7 &	251.1	&	\\
$p=	0.170	$&130.3 &	236.8	&	\\
$p=	0.175	$&131.6 &	230.4	&	\\
\end{tabular}
\end{ruledtabular}
\end{table*}
%

%
\begin{table*}[ht]
\caption{\label{t11} The experimental values of the energy gap amplitude and the $R_{1}$ parameter for Hg1223.}
\begin{ruledtabular}
\begin{tabular}{ccccc}
Hole doping &$T_{C}$ (K)&$\Delta_{tot}^{\left(0\right)}$ (meV)&$R_{1}$&Ref.\\
\hline
$p=0.176$ & 132 & 48 & 8.50 & \cite{Jeong1994A}\\
$p=0.176$ & 132 & 22 & 3.90 &                  \\
\hline
$p=0.157$ & 125   & 24    & 4.5 &  \cite{Rossel1994A}\\
$p=0.181$ & 133   & 38    & 6.6 &                 \\
\end{tabular}
\end{ruledtabular}
\end{table*}
\end{widetext}
\clearpage
%

%
\end{document}